\documentclass[fleqn,twoside]{article}
\usepackage{epsfig,espcrc2}

\usepackage{graphicx}
\usepackage[figuresright]{rotating}

\newcommand{\AmS}{{\protect\the\textfont2
  A\kern-.1667em\lower.5ex\hbox{M}\kern-.125emS}}

\title{The Deconfinement Transition in $SU(4)$ Lattice
Gauge Theory
\vskip-1.8cm\hfill\small hep-lat/0110054, TIFR/TH/01-37\vskip1.5cm
}

\author{Rajiv V. Gavai\address{Department of Theoretical Physics, Tata
                Institute of Fundamental Research, \\
                Homi Bhabha Road, Mumbai 400 005, India}%
        \thanks{E-mail: gavai@tifr.res.in} }
       
\begin{document}

\begin{abstract}
The deconfinement transition in $SU(4)$ lattice gauge theory is studied on
$N_s^3 \times N_t$ lattices with $N_s$ = 8-16 and $N_t$ = 4-8
using a modified Wilson action which is expected to have no bulk
transitions.  The susceptibility $\chi_{|L|}^{\rm max}$ is found to
increase {\it linearly} with spatial volume for $N_t$ = 4, 5, and 6, 
indicating a first order deconfinement phase transition.   The latent
heat is estimated to be $\approx {2 \over 3}$ of the corresponding 
ideal gas energy density at $T_c$.

\vspace{1pc}
\end{abstract}

\maketitle

\section{INTRODUCTION}

The nature of the phase transition(s) to Quark-Gluon Plasma, which we
hope to see in the experiments of RHIC and later of LHC, and the physics 
driving it, have always been of great interest.  While the real world has 
presumably two very light ($u$,$d$) flavours of quarks and one somewhat 
heavier ($s$) flavour, both analytical and numerical methods in lattice 
QCD begin from the limiting cases of either massless or infinitely 
massive quarks.  One talks of the chiral symmetry restoring phase 
transition and the deconfinement phase transition in these two cases 
respectively and has suitable order parameters to investigate them.
For quarks with $N$ colours and $N_f$ massless flavours, these transitions
are related to spontaneous breaking of a global $Z(N)$ and $SU(N_f)
\times SU(N_f)$ chiral symmetry.  Since these symmetries are broken
explicitly to various extents in the real world, which of them is more
relevant is {\it a priori} not clear.  The low masses of the light
flavours suggest chiral symmetry to be the dominant one.  However, it is
seen in numerical simulations that the energy density shows a large
change at the chiral transition and even the order parameter for the deconfinement
transition, 
\begin{equation}
L(\vec x) = {1 \over N} {\rm tr} \prod_{t=1}^{N_t} U_4(\vec x, t)~~,~~
\nonumber
\end{equation}
also rises to nonzero values there.  These apparently mysterious observations 
can be explained\cite{pt} using large $N$ arguments, {\it if} the  
deconfinement transition for $N \ge 4$ is of {\it second order}.  
$SU(4)$ is clearly the simplest case to check this out.

Numerical simulations of $SU(4)$ theory at finite temperatures have been
done in the past\cite{old} and recently\cite{recent} as well. All of
them used the Wilson action, or the more general mixed action:
\begin{equation}
S = \sum_P \big[ \beta \big(1 -{{\rm Re~tr}~ U_P\over N}  \big) +
    \beta_A  \big( 1 - {{\rm tr_A}~ U_P\over N} \big) \big]~,~
\label{mxd}
\end{equation}
A well known problem in the simulations with these actions, especially for 
large $N$, is the presence of a bulk transition which is a lattice artifact. 
The phase diagram of the mixed action,
\begin{figure}[htbp]\begin{center}
\vspace{-0.7cm}
\epsfig{height=5cm,file=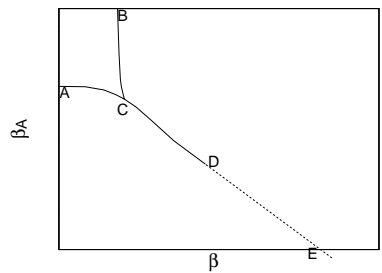, angle=270 }
\vspace{-0.7cm}
\caption{A schematic phase diagram in $(\beta, \beta_A)$-plane for the mixed
action of eq. (\protect{\ref{mxd}}).}
\vspace{-0.7cm}
\label{fg.phd}\end{center}\end{figure}
\vspace{-0.3cm}
in its coupling plane is as shown in Fig. \ref{fg.phd}.  The solid lines in 
it show first order bulk transitions lines. The dotted line after the end 
point D is drawn to suggest the impact D may have on the Wilson axis 
($\beta_A$ = 0).  For $N \ge 4$, D is expected to be where E is shown, 
causing a first order bulk transition for the usual Wilson action.  
In order to avoid it, simulations at negative $\beta_A$ \cite{old} and 
for larger \cite{recent} $N_t$ = 6 were made. They lead to a first order 
deconfinement phase transition for $SU(4)$.

From our extensive studies\cite{us} of the deconfinement phase transition 
for the action above but for the $SU(2)$ theory, we know that bulk 
transitions affect the order and location of the deconfinement transition 
in subtle and inexplicable ways, even leading to apparent qualitative
violations of universality.  Universality was restored \cite{us1} in
that case only after eliminating the bulk transitions associated with 
the $Z(2)$ vortices and $Z(2)$ monopoles by adding large chemical potentials 
for them.  It seems natural to expect that the bulk transitions for $N > 2 $ 
can also be cleaned off by suppressing the corresponding $Z(N)$ objects.  
We pursue this idea here for $SU(4)$ to investigate its deconfinement 
phase transition.

\section{SIMULATIONS AND RESULTS}

Generalizing the idea of positive plaquette models\cite{pp} in the 
literature, we use the action
\begin{equation}
S = \beta \sum_P \big( 1 - {{\rm tr}~ U_P \over N}\big) \cdot \theta(
{\pi \over N} -|\alpha|) ~,~
\end{equation}
where $ -\pi < \alpha \le \pi$ is the phase of tr $U_P$.  By adding the
adjoint term of eq. (2) to the action (3), one sees that the
phase diagram of the resultant mixed should not have the bulk lines AC
or BC and hence the endpoint D or E.

We have simulated the above action on $N_s^3 \times N_t$ lattices for
$N_s$ =8, 10, 12, 15, 16 and $N_t$ = 4, 5, 6, 8 using a 15-hit Metropolis et al.
algorithm. The calculations were done on a cluster of pentiums.
Typically short runs to look for points of rapid variations in
$ \langle |L| \rangle$ were followed by long runs (a few million sweeps)
to determine the susceptibility $\chi_{|L|}$ using the histogramming
technique.  Usual finite size scaling techniques were used to determine 
the order of the transition and its exponents.

\begin{figure}[htbp]\begin{center}
\vspace{-0.7cm}
\epsfig{height=7.5cm,width=6cm,file=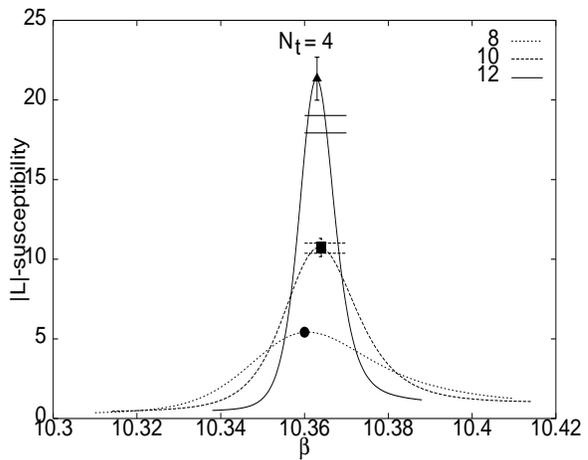,angle=270}
\vspace{-0.7cm}
\caption{ The susceptibility $\chi_{|L|}$ as a function of $\beta$ for
$N_s^3 \times 4$ lattices. }
\vspace{-0.7cm}
\label{fg.susc}\end{center}\end{figure}

In simulations on $N_s^3 \times 4$ lattices, $N_s =$ 8, 10, 12,
one sees hot and cold starts converge quickly at couplings a little 
away from the transition point on its both sides but a clear co-existence of
states is visible for all lattices at the transition point. The
tunneling frequency goes down with spatial volume. The histograms of
$|L|$ show peaks which become narrower with increasing volume and the gap
between them remains unchanged.  These classic signs of a first order phase
transition are confirmed by a quantitative analysis of the linear
growth of $\chi_{|L|}$ with volume, as seen in Fig. \ref{fg.susc}.   
The horizontal lines in each case are predictions obtained by scaling
the $N_s = 8$ results linearly with volume.

\vspace{-0.3cm}

For larger $N_t$, we 
used many longer runs in the region of strong variation of $\langle |L|
\rangle$ to obtain the susceptibility directly and used the histogramming only
for the finer determination of the critical coupling.  
Our results for $\langle |L| \rangle$ as a function of $\beta$ clearly 
show the expected shift for a deconfinement phase transition for $N_t$ 
= 5, 6 and 8.  This is evident in the $\beta_c$ determinations from the 
$\chi_{|L|}^{\rm max}$, as seen in Table 1 for $N_t$ =
6 for two different spatial volumes. Again, using the peak height for the 
smaller spatial volume, the $\chi^{\rm max}$ on the bigger lattice can be 
predicted, assuming a linear growth with volume.  The prediction listed 
in the table can be seen to be in very good agreement with the
direct Monte Carlo determination.  Along with the shifts in $\beta$,
it confirms that the same {\it physical} deconfinement phase
transition is being simulated on these lattices thus approaching the continuum
limit of $a \to 0$ in a progressive manner by keeping the transition
temperature $T_c$ constant in physical units.

\begin{table}
\caption{The values of $\beta$ at which long simulations were performed 
on $N_s^3 \times 6$ lattices, $\beta_c$ and the height of 
the $|L|$-susceptibility peak, $\chi^{\rm max}_{|L|}$.}
\label{table:1}
\medskip
\begin{tabular}{@{}lllll}
\hline
$N_\sigma$ & $\beta$  &  $\beta_{c,N_\sigma}$ & $\chi_{|L|}^{\rm max}$& $\chi^{\rm max}_{\rm predcted}$ \\
\hline
12         &  10.675   &  10.686(5)     &  4.36(35) &   --      \\
16         &  10.675   &  10.676(5) &  10.43(95)&  10.3(8)  \\
\hline
\end{tabular} 
\vspace{-0.7cm}
\end{table}

\section{LATENT HEAT}

While the results above for the deconfinement order parameter $\langle
|L| \rangle$ and the corresponding susceptibility, $\chi_{|L|}$, are
indicative of a first order deconfinement phase transition, one needs to
make sure that they indeed are not due to a coincident first order bulk
transition.  Apart from the characteristic (logarithmic) shift of the 
transition point with $N_t$, seen above, the latent heat of a first
order deconfinement phase transition should also remain constant as 
$N_t \to \infty$.  Requiring the pressure to be
continuous at the deconfinement phase transition, the latent heat can be
obtained from two different observables $ \Delta_1 \equiv 
\Delta (\epsilon-3p)/T_c^4$, and $ \Delta_2 \equiv \Delta (\epsilon+p)/T_c^4$,
where
\begin{eqnarray}
&& \Delta_1  = - 48 N_t^4 a {\partial g^{-2} \over \partial a }
\Delta P \nonumber~,~ \\
&& \Delta_2  = 32 N_t^4 {C(g^2)\over g^2} (\Delta P_t -\Delta
P_s)~,~ 
\label{lat}
\end{eqnarray}
and $C(g^2) = (1 - 0.2366 g^2 + O(g^4))$ for $SU(4)$. $\Delta$ denotes
discontinuities across the transition in the respective variables.  In order
to obtain the $\Delta P$, $\Delta P_s$ and $\Delta P_t$, the minimum of the 
histogram $N(|L|)$ was used to separate the two
phases in each case.  From eq. (\ref{lat}), it is clear that the plaquette
discontinuity $\Delta P \propto N_t^{-4}$ in order to obtain the same latent
heat in physical units, as $N_t \to \infty$.  Indeed, its decrease with 
$N_t$ was seen to be consistent with expectations for $N_t \ge 5$.  
Furthermore, both estimates must agree in this limit, as the neglected 
cut-off corrections become then insignificant. Table
\ref{table:2} suggests this to be the case, leading to an estimate which
is $\approx {2 \over 3}$ of the ideal gas energy density at $T_c$ and 
agrees with earlier results\cite{recent}. 
\begin{table}
\caption{Both the latent heat estimates of eq.(\ref{lat}) as a function of
$N_t$.}
\label{table:2}
\medskip
\begin{tabular}{@{}lllll}
\hline
$N_t$ & 4 & 5 & 6 & 8  \\ 
\hline
$\Delta_1 $&  21.03(5) &  11.02(6) &  8.31(5) &  6.57(16) \\
$\Delta_2$ &  9.89(14) &  7.77(40) &  6.04(60) &  6.45(99)\\
\hline
\end{tabular} 
\vspace{-0.7cm}
\end{table}

\section{SUMMARY}

The linear growth of $\chi_{|L|}^{\rm max}$ with volume for $N_t$ = 4,
suggests a {\it first} order deconfinement phase transition for $SU(4)$. Various
indicators, such as, histograms and evolutions, are in accord with this.
Increasing $N_t$ to 5, 6 and 8 shows the expected shift of the deconfinement 
transition, with a growth in $\chi_{|L|}^{\rm max}$ that is consistent
with being linear in volume. The plaquette discontinuity $\Delta P$ decreases 
as the fourth power of $N_t$, indicating both a lack of a bulk transition and a
first order deconfinement phase transition.  The large estimated latent 
heat, being about 2/3 of the corresponding ideal gas energy density, 
suggests the deconfinement transition to grow stronger in nature as the 
number of colours in increased.

\end{document}